\documentclass[10pt,showpacs,preprintnumbers,amsmath,amssymb]{revtex4}
\usepackage{units}
\usepackage{bbold,mathpazo}%opcoes de fonte: mathptmx, helvet, eulervm, avant, fourier, newcent, mathpazo
\usepackage{graphicx}% Include figure files
\usepackage{dcolumn}% Align table columns on decimal point
\usepackage{bm}% bold math
\usepackage{bbm}
\usepackage{slashed}

\begin{document}

\title{M\"obius transformation for left-derivative quaternion holomorphic functions}
\vspace{2cm}
\author{Sergio Giardino}
\email{p12@ubi.pt}
\affiliation{ Departamento de F\'isica \& Centro de Matem\'atica e Aplica\c c\~oes,  Universidade da Beira Interior\\
Rua Marqu\^es D'\'Avila e Bolama 6200-001 Covilh\~a, Portugal}

\begin{abstract}
\noindent Holomorphic quaternion functions only admit affine functions; thus, the M\"obius transformation for these
functions, which we call quaternionic holomorphic transformation (QHT), only comprises similarity transformations. 
We determine a general group $\mathsf{X}$ which has the group $\mathsf{G}$ of QHT as a particular case. Furthermore, we
observe that the M\"obius group and the Heisenberg group may be obtained by making $\mathsf{X}$ more symmetric.
We provide matrix representations for the group $\mathsf{X}$ and for its algebra $\mathfrak{x}$.
The Lie algebra is neither simple nor semi-simple, and so it is not classified among the classical 
Lie algebras. They prove that the group $\mathsf{G}$ comprises $\mathsf{SU}(2,\mathbb{C})$ rotations, dilations and
translations. The only fixed point of the QHT is located at infinity, and the 
QHT does not admit a cross-ratio. Physical applications are addressed at the conclusion.
\end{abstract}

\maketitle

\section{Introduction\label{S1}}
For complex $a, b, c, d$ and $z$, the M\"obius transformation (MT) is defined by 
\begin{equation}\label{moebius}
M(z)=\frac{az+b}{cz+d}.
\end{equation}
This expression may be decomposed into the symmetry operations of translation, rotation, dilation and the special conformal 
transformation, and a $\mathsf{SL}(2,\,\mathbb{C})$ matrix group representation is naturally built as well. The
M\"obius transformation is also the symmetry of hyperbolic geometry \cite{Anderson:2008geo} and the symmetry it contains has a broad application in physics
through the conformal field theory \cite{Ginsparg:1988ui,Blumenhagen:2009zz}. The generalization of MT for higher dimensional 
flat geometries is also a well-known subject 
\cite{Wilker:1993,Waterman:1993msd,Beardon:1995gdg,Porteous:1995acg,Lounesto:1997cas,Jeromin:2003mdg}.

On the other hand, (\ref{moebius}) may be generalized using hyper-complex numbers. A quaternionic M\"obius transformation
\cite{Sudbery:1979qta,Short:2007qmt,Lounesto:1997cas} may be defined according to 
\begin{equation}\label{quat_mob}
\mu(q)=(aq+b)(cq+d)^{-1},\qquad\mbox{with}\qquad \Delta=|a|^2|d|^2+|b|^2|c|^2-2\Re[a\bar cd\bar b]\neq 0
\end{equation}
and quaternionic $a,\,b,\,c,\,d,$ and $q$. The non-commutativity of quaternions implies that the transfomation (\ref{quat_mob})
 is not only a M\"obius group in four dimensions, and even a matrix representation is not natural,
due to quaternion matrices being more restrictive than complex matrices \cite{Sprossig:1998qcc}.
Nevertheless, there is new research in the field, such as attempts to classify the quaternionic M\"obius transformations
\cite{Short:2007qmt,Short:2011ccq}, their properties \cite{Huang:2013dst}, and transformations involving alternative definitions 
for quaternion functions \cite{Stoppato:2011rmt}.

These previous studies consider quaternion M\"obius transformations comprising all symmetry operations found in 
two-dimensional
transformations. The approach used in this article is different because the quaternion transformations are defined for 
holomorphic quaternionic functions. This class of functions only admits affine quaternion functions, given by
\begin{equation}\label{aqf}
\mathcal{F}(q)=qa+b,
\end{equation}
with quaternionic $a, b$ and $q$. The order of the product $qa$ is fixed because of quaternion 
non-commutativity. In order to investigate a M\"obius transformation, we determine infinitesimal Lie 
operators that transform (\ref{aqf}),
and we determine a matrix group representations for the algebra and for the group as well. We call this matrix
representation $\mathsf{X}$, and its algebra is $\mathfrak{x}$.
Assuming suitable constraints, the 
Lie group $\mathsf{X}$ may be identified with the usual M\"obius group $\mathsf{M}$, the Heisenberg group $\mathsf{H}$ and the 
group $\mathsf{G}$ of quaternion holomorphic
transformation (QHT) of (\ref{aqf}). Accordingly, the group $\mathsf{X}$ generalizes these groups, and this 
generalization permits us to unify the usual differential operators 
of the M\"obius algebra, the Heisenberg algebra and  the QHT algebra into a single algebra, $\mathfrak{x}$. Additionally
we determine that the QHT group $\mathsf{G}$ comprises $\mathsf{SU}(2,\mathbb{C})$ rotations, dilations and translations, 
as expected.

The article is organized as follows; in Section \ref{S2}, we give a brief survey of quaternion holomorphic functions
and define the quaternion holomorphic transformation. In section \ref{S3}, we determine the Lie algebra $\mathfrak{x}$ in
terms of differential Lie operators and provide matrix representations for the algebra and for  
the Lie group $\mathsf{X}$. In section \ref{S4}, we study several properties of the QHT and relate them to $\mathsf{X}$. Furthermore,
we relate this group to the M\"obius group, the Heisenberg group and the QHT group.  Section \ref{S5} presents
the Lie algebra $\mathfrak{g}$ of $\mathsf{G}$ in detail. Section \ref{S6} presents our conclusions and future perspectives.

\section{functions and transformations\label{S2}}
\subsection{Holomorphic quaternion functions}
In this section, we give a quaternion analogue for the Cauchy-Riemann condition for complex functions. Further details on quaternion analysis
may be found at \cite{Sudbery:1979qta,Deavours:1973qtc}. We begin defining the left-derivative of a quaternion valued 
function $\mathcal{F}$, so that
\begin{equation}
\frac{d\mathcal{F}}{dq}=\lim_{\Delta\to 0}\frac{1}{\Delta}\big[\mathcal{F}(q+\Delta)-\mathcal{F}(q)\big],
\end{equation}
where $q$ is the quaternionic variable. Remembering quaternionic non-commutativity, the differential relation for the left derivative is defined in the product
\begin{equation}\label{difquat}
 d\mathcal{F}=dq\frac{d\mathcal{F}}{dq}.
\end{equation}
Using the extended notation for quaternions in terms of real variables $x_0$, $x_1$, $x_2$, and $x_3$, we get
\begin{equation}
q=x_0+x_1i+x_2j+x_3k, 
\end{equation}
and then the differential relation (\ref{difquat}) becomes
\begin{equation} 
\partial_0\mathcal{F}\,dx_0+\partial_1\mathcal{F}\,dx_1+\partial_2\mathcal{F}\,dx_2+\partial_3\mathcal{F}\,dx_3=
(dx_0+dx_1i+dx_2j+dx_3k)\frac{d\mathcal{F}}{dq}.
\end{equation}
Equating the real differential coefficients, we get
\begin{equation}\label{difreal}
\frac{d\mathcal{F}}{dq}=\partial_0\mathcal{F}=-i\,\partial_1\mathcal{F}=-j\,\partial_2\mathcal{F}=-k\,\partial_3\mathcal{F}.
\end{equation}
We now define the symplectic notation, where quaternions are written as 
\begin{equation}\label{q_simpletica}
q=z+\zeta j
\end{equation}
where both $z$ and $\zeta$ are complex. Accordingly, a quaternionic
function is written in symplectic notation as
\begin{equation}\label{F_simpletica}
\mathcal{F}=\mathcal{E}_0+\mathcal{E}_1j,
\end{equation}
with $\mathcal{E}_0$ and $\mathcal{E}_1$ complex functions. From (\ref{difreal}) and (\ref{F_simpletica}), we
obtain
\begin{equation}\label{CR_pre}
\partial_0 \mathcal{E}_0=-i\partial_1 \mathcal{E}_0=\partial_2\bar{\mathcal{E}}_1=i\partial_3 \bar{\mathcal{E}_1},
\qquad\partial_0 \bar{\mathcal{E}_1}=i\partial_1 \bar{\mathcal{E}_1}=-\partial_2 \mathcal{E}_0=i\partial_3 \mathcal{E}_0
\qquad\mbox{and}\qquad \partial_{\bar z}\mathcal{F}=\partial_{\bar \zeta} \mathcal{F}=0,
\end{equation}
where $z=x_0+ix_1$, $\zeta=x_2+ix_3$ and  
\[
 \partial_z=\partial_0-i\partial_1\qquad\mbox{ and}\qquad \partial_\zeta=\partial_2-i\partial_3.
\]
From (\ref{CR_pre}) we obtain 
\begin{equation}
\partial_z \mathcal{E}_0=\partial_{\bar\zeta}\bar{\mathcal{E}_1}\qquad\mbox{and}\qquad\partial_\zeta \mathcal{E}_0=-\partial_{\bar z}\bar{\mathcal{E}_1}.
\end{equation}
The above results permit us to state that $\mathcal{F}$ does not depend on the complex conjugate quaternion variable $\bar q$.
The analysis of the second derivatives, which may be found in  \cite{Sudbery:1979qta}, permits us to conclude that the 
only non-trivial left derivative quaternionic function is the affine function, so that
\begin{equation}\label{qaffine}
\mathcal{F}=q\,a+b,
\end{equation}
with $a$ and $b$ quaternion constants. The affine quaternionic function (\ref{qaffine}) is what we define as a holomorphic
quaternion function. It is essential to note that (\ref{qaffine}) do not satisfy a quaternionic Cauchy
theorem, where $\oint dzf(z)=0$. A Cauchy-Riemann theorem for quaternionic function is valid only for regular quaternionic
functions \cite{Sudbery:1979qta,Deavours:1973qtc}, and (\ref{qaffine}) does not belong to this class. Regular quaternionic 
functions may be used to build the quaternionic M\"obius transformation (\ref{quat_mob}), something different that is being 
done here, where only quaternion holomorphic functions are used. Only for completeness, using $\mathcal{F}(\bar q)=\mathcal E_0+\mathcal E_1j$ we have
\begin{eqnarray}
&&\partial_0 \mathcal{E}_0=i\partial_1 \mathcal{E}_0=-\partial_2\bar{\mathcal{E}}_1=-i\partial_3 \bar{\mathcal{E}_1},
\qquad\partial_0 \bar{\mathcal{E}_1}=-i\partial_1 \bar{\mathcal{E}_1}=\partial_2 \mathcal{E}_0=-i\partial_3 \mathcal{E}_0\\
&&\qquad \partial_{ z}\mathcal{F}=\partial_{\bar \zeta} \mathcal{F}=0\qquad
 \partial_{\bar z} \mathcal{E}_0=-\partial_{\bar\zeta}\bar{\mathcal{E}_1}\qquad\mbox{and}\qquad\partial_\zeta \mathcal{E}_0=\partial_{z}\bar{\mathcal{E}_1},
\end{eqnarray}
and consequently $\mathcal F=\bar{q} a+b$.
\subsection{The quaternionic holomorphic transformation (QHT)}
A quaternionic holomorphic transformation (QHT) is presumed to relate to
holomorphic functions. This transformation cannot be a quaternionic  analogue of (\ref{moebius}) like (\ref{quat_mob})
because these 
transformations involve the inversion of quaternions, and the inverse of a quaternionic variable is not a holomorphic
quaternion function. Accordingly, the QHT is another affine quaternion function, so that
\begin{equation}\label{qht}
\mathnormal{G}(q)=q\,u+v.
\end{equation}
The QHT only comprises  a translation given by $v$ and a rotation and a dilation included in $u$, and not the inversion symmetry 
operation. The (\ref{qht}) transformation is thus  classified as a similarity transformation.

\section{Lie algebra and Lie group of the QHT\label{S3}}
As discussed in the previous section, the M\"obius transformation in two dimensions has four symmetry operations,
the translation, namely the rotation, the dilation and the special conformal transformation. Left-derivative quaternion functions
exclude the special conformal transformation, because they contain the inversion operation, which generate a non-analytic quaternion
 function. Thus, the quaternionic transformation we consider generates linear functions like (\ref{qaffine}).
Multiplication of the quaternionic variable by a quaternion constant comprises rotations and dilations and the addition
of a quaternionic constant to a quaternion variable comprises translations. Let us then consider an infinitesimal similarity
transformation
\begin{equation}
 q\to q(1+\epsilon)+\delta
\end{equation}
where $\epsilon$ and $\delta$ are infinitesimal quaternionic constants. Using the aforementioned transformation on the quaternion
function (\ref{qaffine}) we obtain
\begin{equation}\label{inf_transf}
 \mathcal{F}\to [q(1+\epsilon)+\delta]a+b=\mathcal{F}+q\epsilon a+\delta a.
\end{equation}
The generator of the translation may simply be $\partial_q$, but the differential generator for the transformation
involving $\epsilon$ cannot be $q\partial_q$. Because of the non-commutativity of quaternions, the $\epsilon$ may not
be put on the left of $q$ in order to get a differential generator for $q\epsilon a$ in (\ref{inf_transf}). In order
to get the Lie algebra generators, we then consider the quaternion generator comprising two complex parts. Thus, using
symplectic notation, where
\[
 \epsilon=\epsilon_0+\epsilon_1j
\]
for complex $\epsilon_0$ and $\epsilon_1$, we write  
\begin{eqnarray}
q\epsilon a&=&(z+\zeta j)(\epsilon_0+\epsilon_1j)a\nonumber\\
&=&\big[\epsilon_0z-\bar\epsilon_1\zeta+(\epsilon_1z+\bar\epsilon_0\zeta)j\big]a\nonumber\\
&=&(\epsilon_0z-\bar\epsilon_1\zeta)\partial_z\mathcal{F}+(\epsilon_1z+\bar\epsilon_0\zeta)\partial_\zeta\mathcal{F}
\end{eqnarray}
Then, we have four generators: $z\partial_z$,  $z\partial_\zeta$,
 $\zeta\partial_z$ and  $\zeta\partial_\zeta$. Furthermore the translation represented by the third term of (\ref{inf_transf})
is generated by $\partial_z$ and  $\partial_\zeta$, and hence we have a total of six generators . On the other hand, we 
repeat the procedure for the complex conjugated quaternion variable $\bar q$ on a left-derivative holomorphic function $\mathcal{G}$,
and hence we obtain
\begin{equation}
\mathcal{G}\to [\bar q(1+\epsilon)+\delta]a+b=\mathcal{G}+\bar q\epsilon a+\delta a. 
\end{equation}
Consequently, from $\bar q=\bar z -\zeta j$ we obtain
\begin{equation}
\bar q\epsilon=(\bar z-\zeta j)(\epsilon_0+\epsilon_1j)=\epsilon_0\bar z+\bar\epsilon_1\zeta+(\epsilon_1\bar z-\bar\epsilon_0\zeta)j.
\end{equation}
Consequently, $\bar z\partial_{\bar z}$,  $\bar z\partial_\zeta$,
 $\zeta\partial_{\bar z}$ and  $\zeta\partial_\zeta$ are the generators for the dilations and for the rotations,
 and $\partial_{\bar z}$ and  $\partial_\zeta$ are the generators
for the translation. Note that the function $\mathcal{G}$ cannot be obtained from a complex conjugation of a left-derivatives
holomorphic function, because $\bar{\mathcal{F}}=\bar a\bar q+\bar b$ is not left-derivative.

Naming the generator sets of the Lie algebras for the quaternionic transformations as $\mathfrak{x}=\{x_i\}$
for a quaternionic variable and as $\bar{\mathfrak{x}}=\{\bar x_i\}$ for a complex conjugated variable, we get
\begin{eqnarray}
&& x_1=\partial_z,\qquad x_2=z\partial_z\qquad x_3=\zeta\partial_z \qquad x_4=\partial_\zeta,  \qquad x_5=\zeta\partial_\zeta,
\qquad x_6 =z\partial_\zeta,\qquad\mbox{and}\nonumber\\
&& \bar x_1=\partial_{\bar z},\qquad \bar x_2=\bar z\partial_{\bar z}\qquad \bar x_3=\zeta\partial_{\bar z} \qquad 
\bar x_4=\partial_\zeta,  \qquad \bar x_5=\zeta\partial_\zeta, \qquad \bar x_6 =\bar z\partial_\zeta.\label{x-alg}
\end{eqnarray}
We note that the generators $\partial_\zeta$ and $\zeta\partial_\zeta$ are common for both 
the cases. The Lie algebra for $\mathfrak{x}$ is written through the commutation relations
\begin{eqnarray}
&&\nonumber [x_1,\,x_2]=x_1\\
&&\nonumber [x_1,\,x_3]=0\qquad [x_2,\,x_3]=-x_3\\
&&\nonumber [x_1,\,x_4]=0\qquad [x_2,\,x_4]=0 \qquad [x_3,\,x_4]=-x_1\\
&&\nonumber [x_1,\,x_5]=0\qquad [x_2,\,x_5]=0\qquad [x_3,\,x_5]=-x_3,\qquad [x_4,\,x_5]=x_4\\
&&\nonumber [x_1,\,x_6]=x_4\;\;\;\;\,\, [x_2,\,x_6]=x_6\;\;\;\;\;\, [x_3,\,x_6]=x_5-x_2\;\;\;\,\, [x_4,\,x_6]=0\qquad [x_5,\,x_6]=-x_6
\end{eqnarray}
As we will see below, the $\mathfrak{x}$ algebra is not the QHT algebra, but we will need to adopt a constraint
for recovering the QHT from $\mathfrak{x}$. The algebra $\bar{\mathfrak{x}}$ is of course isomorphic to $\mathfrak{x}$. 

On the other hand, the direct sum of 
$\mathfrak{x}$ and $\bar{\mathfrak{x}}$ is not an algebra. Due to the commutator
\begin{equation}
[x_6,\,\bar x_3]=[z\partial_\zeta,\,\zeta\partial_{\bar z}]=z\partial_{\bar z},
\end{equation}
$\mathfrak{x}\oplus \bar{\mathfrak{x}}$ is not closed, and then these algebras must always be considered separately. Perhaps this
direct sum may be achieved using a right-derivative holomorphic function, but this hypothesis has not been considered here.

We can obtain the adjoint matrix representation for this algebra using the structure constants, so that the matrix elements 
could be obtained from
$(x_i)_j^{\;k}=c_{ij}^{\;\;\;k}$. However, this representation comprises sixth order matrices that are cumbersome
to manipulate. A simpler representation may be obtained defining a linear space represented by  the vector 
\begin{equation}\label{vec}
\bm{v}=(z,\,\zeta,\,1).
\end{equation}
Let us then act on $\bm{v}$  the differential generators (\ref{x-alg}) of $\mathfrak{x}$ in order to determine a matrix representation of the algebra.
Collecting the results for each generator and adjusting the signals, we obtain a third order matrices representation
\begin{eqnarray}\label{xgenerators}
\mathsf{x}_1=\left[
\begin{array}{ccc}
0\; 0\; 1\\
0\; 0\; 0\\
0\; 0\; 0
\end{array}
\right]\qquad
& \mathsf{x}_2=(-1)\left[
\begin{array}{rrr}
1\; 0\; 0\\
0\; 0\; 0\\
0\; 0\; 0
\end{array}
\right] &\qquad 
\mathsf{x}_3=\left[
\begin{array}{ccc}
0\; 1\; 0\\
0\; 0\; 0\\
0\; 0\; 0
\end{array}
\right]\nonumber\\
\mathsf{x}_4=(-1)\left[
\begin{array}{ccc}
0\; 0\; 0\\
0\; 0\; 1\\
0\; 0\; 0
\end{array}
\right]\qquad
& \mathsf{x}_5=(-1)\left[
\begin{array}{rrr}
0\; 0\; 0\\
0\; 1\; 0\\
0\; 0\; 0
\end{array}
\right] &\qquad 
\mathsf{x}_6=\left[
\begin{array}{ccc}
0\; 0\; 0\\
1\; 0\; 0\\
0\; 0\; 0
\end{array}
\right].
\end{eqnarray}
From the commutation relations of the algebra, we can see that $\mathfrak{x}$  has a non-trivial  ideal $I$, so that 
$I=\{x_1,\,x_4\}$, and thus $\mathfrak{x}$ is neither a simple nor a semi-simple algebra.
 We may get further understanding looking at the Lie group, which may be built using the exponential mapping. 
If $\mathsf{X}_i$ is an element of the Lie matrix group $\mathsf{X}$, which is the covering Lie group of $\mathfrak{x}$, then
\begin{equation}
\mathsf{X}_i=\exp(tx_i)=\left\{
\begin{array}{lr}
\mathbb{1}+t\,\mathsf{x}_i \qquad\qquad\qquad\mbox{for}\qquad i=\{1,\,3,\,4,\,6\} \\ 
\mathbb{1}+(1-e^{-t})\mathsf{x}_i \qquad\;\;\mbox{for}\qquad i=\{2,\,5\}
\end{array}
\right.
\end{equation}
so that $t$ is a real parameter. This group can be represented by the matrix
\begin{equation}\label{xmatrix}
\mathsf{X}=\left[\begin{array}{ccc}
a\; b\; p\\
c\; d\; q\\
0\; 0\; 1
\end{array}\right],
\end{equation}
where $\mathsf{X}$ has complex entries. We willl now examine several properties of the Lie group represented by the $\mathsf{X}$
 matrix in order to illustrate its relation to the QHT (\ref{qht}).
\section{Properties of the Lie group $\mathsf{X}$\label{S4}}
First, we shall define a map $m:V\to\mathbb{H}$ between the vector space $V$ which contains $\bm{v}$ defined in
(\ref{vec}) and a quaternion number $q=z+\zeta j$, so that
\begin{equation}
 m(\bm{v}) =q.
\end{equation}
Thence we use $m$ to relate $\mathsf{X}\bm{v}$ to the QHT. In general,
\begin{equation}
\mathsf{X}\bm{v}=\left[\begin{array}{ccc}
a\; b\; p\\
c\; d\; q\\
0\; 0\; 1
\end{array}\right]
\left[
\begin{array}{c}
z\\ \zeta\\1       
\end{array}
\right]=
\left[
\begin{array}{c}
az+b\zeta+p\\ cz+d\zeta+q\\1       
\end{array}
\right],\qquad\mbox{and}\qquad
m(\mathsf{X}\bm{v})=az+b\zeta+p+(cz+d\zeta+q)j.
\end{equation}
The $a, b, c$ and $d$ entries promote rotations and dilations on $z$ and $\zeta$, and $p$ and $q$
generate quaternionic translations. The group $\mathsf{X}$ defines a transformation on $q$ that is more
general than a QHT. In order to obtain a QHT (\ref{qht}) we must impose several constraints on $\mathsf{X}$. We shall then 
write a general QHT as
\begin{eqnarray}
\mathnormal{G}(q)&=&q\,u+v\nonumber\\
&=&(z+\zeta j)(a+bj)+p+qj\nonumber\\
&=&az-\bar b\zeta+(b z+\bar a \zeta)j+p+qj.
\end{eqnarray}
A QHT may then be represented by a matrix $\mathsf{G}$, so that
\begin{equation}\label{qht_matrix}
\mathnormal{G}(q)\Leftrightarrow m\left(\mathsf{G} \bm{v}\right), \qquad\mbox{for}\qquad 
\mathsf{G}=\left[\begin{array}{lll}
a\; -\bar b\;\; p\\
b\; \;\;\;\;\bar a\;\; q\\
0\; \;\;\;\;0\;\; 1
\end{array}\right].
\end{equation}
Then we have obtained the group $\mathsf{G}$ of QHT as a restriction of the group $\mathsf{X}$.
The determinant of $\mathsf{X}$, given by $|X|=ad-bc$ may have an arbitrary value. After the discussion of the QHT cross 
ratio and the fixed points in the sequel, we will study the consequences of normalizing this determinant.
\subsection{Cross-ratio and fixed points}
A general M\"obius transformation (\ref{moebius}) admits the existence of an invariant relation. This function
is the cross-ratio, given by the expression
\begin{equation}\label{cross}
 [z,\,z_1,\,z_2,\,z_3]=\frac{(z-z_1)(z_2-z_3)}{(z-z_3)(z_2-z_1)}.
\end{equation}
Using the identity
\begin{equation}
 M(z)-M(z_1)=\frac{(ad-bc)(z-z_1)}{(cz+d)(cz_1+d)},
\end{equation}
the cross-ratio may be easily obtained. (\ref{cross})  has the important property of invariance under M\"obius transformations,
and then
\begin{equation}
 [z,\,z_1,\,z_2,\,z_3]=[M(z),\,M(z_1),\,M(z_2),\,M(z_3)].
\end{equation}
The cross-ratio implies, if an arbitrary M\"obius transformation $M$ is applied on three of any four points, 
$M$ is the unique transformation that, applied on the fourth, will preserve the cross-ratio built with the other
three transformed points. Accordingly, it can be proven
that three arbitrary points are related to another set of three points by an unique M\"obius transformation. A standard
three points set comprises $0$, $1$, and $\infty$, and then there is a unique M\"obius transformation that relates this
set to an arbitrary set of three complex numbers.

In order to determine whether there is some kind of cross ratio for the QHT, we could take the complex similarity transformation
\begin{equation}\label{simil}
N(z)=az+b,
\end{equation}
and obtain a cross ratio using
\[
N(z)-N(z_1)=(z-z_1)a, \qquad\mbox{so that}\qquad [z,\,z_1,\,z_2]=\frac{z-z_1}{z-z_2}.
\]
Considering complex numbers as vectors, the similarity transformation (\ref{simil}) rotates both of the vectors by the same angle 
and dilates their norms by the same factor, and consequently the difference vector suffers the same transformation. The cross-ratio,
on this side means that, for any given two points, a third point and its image are related by a unique similarity condition. 
On the other hand, for the QHT, the relation
\begin{equation}\label{qht_dif}
G(q)-G(q_1)=(q-q_1)(a+bj)
\end{equation}
informs us that the rotation and dilation are preserved in the difference. However, due to the anti-commutativity of quaternions, 
we cannot build a cross ratio to prove that there is a unique QHT relating every pair of quaternions.
This is the first indication of a lack of symmetry of the QHT compared to the complex M\"obius transformation.
 In the next section we discuss constraints that lead QHT to more symmetric cases. The difference (\ref{qht_dif}),
may generate a cross ratio by restricting the QHT to translations or real dilations, for example.

The QHT, as the similarity transformation for complex numbers, has only one fixed point, which obeys
\[
 G(q)=q.
\]
For the complex case, a M\"obius transformation with only one fixed point is classified as belonging to the parabolic type. 
This kind of transformation has a fixed point at infinity, and this fixed point cannot be moved to a finite coordinate complex
point by a parabolic
transformation. In a general M\"obius transformation, due to the inverse operation, the infinity may be  related to the zero point
by inversion. The QHT, considered a parabolic transformation on each complex number that constitutes the quaternion
in the symplectic notation, has analogously the infinity as its only fixed point, which cannot be moved to another point
by any QHT, and consequently infinity is the only fixed point of the QHT.

\subsection{$\mathsf{X}$ group as a M\"obius transformation\label{xMt}}
Imposing $p=q=0$ on (\ref{xmatrix}), we have a matricial representation of the M\"obius group, represented by 
\begin{equation}\label{xmobius}
\mathsf{M}=\left[\begin{array}{lll}
a\; b\; 0\\
c\; d\; 0\\
0\; 0\; 1
\end{array}\right].
\end{equation}
In terms of quaternions, this representation is simply a translationless transformation
\begin{equation}
m(\mathsf{M}\bm{v})=az+b\zeta+(cz+d\zeta)j.
\end{equation}
In terms of differential operators, the algebra of the group is given by the set 
subset $\{x_2,\,x_3,\,x_5,\,x_6\}$ of (\ref{x-alg}). 
On the other hand, by imposing $|\mathsf{M}|=1$, we have the $\mathsf{sl}(2,\,\mathbb{C})$
algebra of the normalized M\"obius group. This is achieved by using the operators set comprising 
\begin{equation}\label{qho}
\{x_3,\,x_6,\,x_5-x_2\}=\left\{\zeta\partial_z,\,z\partial_\zeta,\, \zeta\partial_\zeta-z\partial_z\right\},
\end{equation}
whose elements satisfy the algebra
\begin{equation}
 [x_2-x_5,\,x_3]=[x_2-x_5,\,x_6]=0,\qquad[x_3,\,x_6]=x_5-x_2.
\end{equation}
Then we have a representation of the $\mathsf{SL}(2,\,\mathbb{C})$ group, as expected. The representation by the differential
operators (\ref{xmobius}) is not new, but its interpretation in terms of quaternions is certainly new. In terms of physics,
it is fascinating to observe that the above operators may be used to represent the quantum harmonic oscillator.

\subsection{$\mathsf{X}$ group as the Heisenberg group}
The Heisenberg group, also called the Weyl group, comprises upper triangular matrices of the form
\begin{equation}\label{h1_matrix}
\mathsf{H}=\left[\begin{array}{lll}
1\; b\; p\\
0\; 1\; q\\
0\; 0\; 1
\end{array}\right].
\end{equation}
The group is isomorphic to the group of matrix of the type
\begin{equation}\label{h2_matrix}
\tilde{\mathsf{H}}=\left[\begin{array}{lll}
1\; 0\; p\\
c\; 1\; q\\
0\; 0\; 1
\end{array}\right].
\end{equation}
Thus algebra of the group may be represented by matrices either as
\begin{equation}
\mathsf{h}=\left[\begin{array}{lll}
0\; b\; p\\
0\; 0\; q\\
0\; 0\; 0
\end{array}\right]
\qquad\mbox{or as}\qquad
\tilde{\mathsf{h}}=\left[\begin{array}{lll}
0\; 0\; p\\
c\; 0\; q\\
0\; 0\; 0
\end{array}\right].
\end{equation}
These algebras may be represented by sub-algebras of $\mathsf{X}$, either the matrix generators 
$\{\mathsf{x}_1,\,\mathsf{x}_4,\,\mathsf{x}_3\}$ or $\{\mathsf{x}_1,\,\mathsf{x}_4,\,\mathsf{x}_6\}$. The Heisenberg
algebra is then a sub-algebra of $\mathfrak{x}$, but it is not a sub-algebra of the algebra of a QHT. Only
for $b=0$ in $\mathsf{h}$ and for $c=0$ in $\tilde{\mathsf{h}}$ the algebras coincide. In this sense, only
the translation of a QHT is identified with a sub-algebra of Heisenberg.

In terms of differential operators, the Heisenberg algebra is  usually represented as
\begin{equation}
 X=\partial_x-\frac{1}{2}y\partial_z,\qquad Y=\partial_y+\frac{1}{2}x\partial_z,\qquad Z=\partial_z,
\end{equation}
which obey the algebra
\begin{equation}
 [X,\,Y]=Z,\qquad[Z,\,X]=[Z,\,Y]=0 .
\end{equation}
The operators $X$ and $Y$ may be interpreted as momentum and position operators in certain quantum mechanical applications
\cite{Binz:2008ghg}. This algebra may be represented by two subsets of (\ref{x-alg}), either by 
\begin{equation} 
\{x_1,\,x_4,\,x_3\}=\left\{\partial_z,\,\partial_\zeta,\,z\partial_z\right\} \qquad\mbox{or by}\qquad
\{x_1,\,x_4,\,x_6\}=\left\{\partial_z,\,\partial_\zeta,\,\zeta\partial_\zeta\right\}.
\end{equation}
Then, in terms of the differential operators $x_i$ of $\mathfrak{x}$, we have a new representation for the Heisenberg algebra.

\section{The QHT Lie algebra\label{S5}}
The QHT group was obtained as represented by the matrix (\ref{qht_matrix})
\begin{equation}
\mathsf{G}=\left[\begin{array}{lll}
a\; -\bar b\;\; p\\
b\; \;\;\;\;\bar a\;\; q\\
0\; \;\;\;\;0\;\; 1
\end{array}\right].
\end{equation}
The set of matrix generators for its algebra $\mathfrak{g}$ may be obtained from the $\mathsf{x}_i$ generators of 
$\mathfrak{x}$ (\ref{xgenerators}) defining
\begin{equation}\label{xg}
\mathsf{g}_1=\mathsf{x}_3+\mathsf{x}_6,\qquad \mathsf{g}_2=i(\mathsf{x}_6-\mathsf{x}_3),\qquad\mathsf{g}_3=\mathsf{x}_2-\mathsf{x}_5,\qquad
\mathsf{g}_4=-(\mathsf{x}_2+\mathsf{x}_5),\qquad \mathsf{g}_5=\mathsf{x}_1,\qquad\mathsf{g}_6=-\mathsf{x}_4,
\end{equation}
so that
\begin{eqnarray}\label{ggenerators}
\mathsf{g}_1=\left[
\begin{array}{lll}
0\; 1\; 0\\
1\; 0\; 0\\
0\; 0\; 0
\end{array}
\right]\qquad
& \mathsf{g}_2=\left[
\begin{array}{lll}
0\;-i\;\; 0\\
i\;\;\;\;\;\, 0\;\; 0\\
0\;\;\;\;\, 0\;\; 0
\end{array}
\right] &\qquad 
\mathsf{g}_3=\left[
\begin{array}{lll}
1\;\;\;\;\; 0\;\;\, 0\\
0\; -1\;\; 0\\
0\;\;\;\;\; 0\;\;\, 0
\end{array}
\right]\nonumber\\
\mathsf{g}_4=\left[
\begin{array}{ccc}
1\; 0\; 0\\
0\; 1\; 0\\
0\; 0\; 0
\end{array}
\right]\qquad
& \mathsf{g}_5=\left[
\begin{array}{rrr}
0\; 0\; 1\\
0\; 0\; 0\\
0\; 0\; 0
\end{array}
\right] &\qquad 
\mathsf{g}_6=\left[
\begin{array}{ccc}
0\; 0\; 0\\
0\; 0\; 1\\
0\; 0\; 0
\end{array}
\right].
\end{eqnarray}
From (\ref{xg}) and from (\ref{x-alg}), we obtain the representation in terms of the differential operators
\begin{eqnarray}
&g_1=\zeta\partial_z+z\partial_\zeta,\qquad &g_2=i(z\partial_\zeta-\zeta\partial_z),\qquad g_3=z\partial_z-\zeta\partial_\zeta,\\
\;&g_4=-z\partial_z-\zeta\partial_\zeta,\qquad &g_5=\partial_z,\qquad\qquad\qquad\; g_6=-\partial_\zeta,\nonumber
\end{eqnarray}
which obeys the algebra
\begin{eqnarray}
&&\nonumber [g_1,\,g_2]=2ig_3\\
&&\nonumber [g_1,\,g_3]=2ig_2\;\;\;\;[g_2,\,g_3]=2ig_1\\
&&\nonumber [g_1,\,g_4]=0\qquad\,\; [g_2,\,g_4]=0 \qquad\;\;\; [g_3,\,g_4]=0\\
&&\nonumber [g_1,\,g_5]=g_6\qquad [g_2,\,g_5]=ig_6\qquad [g_3,\,g_5]=-g_5,\;\;\;[g_4,\,g_5]=g_5\\
&&\nonumber [g_1,\,g_6]=g_5\qquad [g_2,\,g_6]=-ig_5\;\;\;\; [g_3,\,g_6]=g_6\qquad\, [g_4,\,g_6]=g_6\qquad [g_5,\,g_6]=0
\end{eqnarray}
We see that the sub-algebra $\{g_1,\,g_2,\,g_3\}$ generates the $\mathsf{su}(2,\,\mathbb{C})$ algebra, and then we can
associate these operators with rotations on a two sphere. The operator $g_4$ is responsible for dilations. If we exclude it
from $\mathfrak{g}$, this is equivalent to impose  $|\mathsf{G}|=1$. This phenomenon has already been observed in the 
discussion of $\mathsf{X}$ as a M\"obius group in Section \ref{xMt}.

\section{conclusion\label{S6}}
In this article we have presented the Lie algebra for a similarity transformation of quaternion holomorphic functions. 
The similarity transformation is the M\"obius transformation for quaternion holomorphic functions, which does not have the inversion
operation because this would generate a non-holomorphic quaternion fuction. 

For this purpose, we studied the group of symmetry of a transformation $X$, in which independent similarity 
transformations are applied to the complex components of a quaternion, and we have called this group $\mathsf{X}$.
Compared to the complex M\"obius transformation $M$, 
quaternionic transformation $X$ is less symmetric and admits further possibilities. Adopting some constraints on 
 $\mathsf{X}$, we obtain the complex M\"obius group $\mathsf{M}$, the Heisenberg group $\mathsf{H}$ and the
QHT group $\mathsf{G}$. Thus, we have set very symmetric groups as particular cases of a group $\mathsf{X}$ with
 many degrees of freedom.
From the mathematical standpoint, a possible future direction for research is to investigate
 whether the group presented here may be used for geometric studies, as has been done for conformal differential
geometry \cite{Jeromin:2003mdg}. The study of right-derivative holomorphic functions also seems to be an interesting 
direction for research.

There may be several physical applications for $\mathsf{X}$ and $\mathsf{G}$, and we have just mentioned a several few among
many others we are unable to imagine. Some of them are very simple, and this makes these  possibilities interesting. 
The quantum harmonic oscillator could be studied through the operators (\ref{qho}), and a new connection between
complex quantum mechanics and quaternionic quantum mechanics \cite{Adler:1995qqm} may be established. Also the study of quaternonic
quantum field theory may be benefited \cite{Giardino:2012ti}.
As the conformal symmetry is present in $\mathsf{G}$, the study of quantum 
mechanics \cite{deAlfaro:1976je} and field theories \cite{Fubini:1976jm}
using conformal symmetry may also be studied using the algebra presented here, and the connection to quaternion
quantum mechanics might also be tested here as well. Complex numbers have been used to build the two-dimensional conformal field theory, and quaternions may
be used to build a four-dimensional conformal field theory. In fact, there are already some attempts in this direction,
using quaternionic matrices, quaternionic projective spaces and quaternionic Fuller analytic functions
\cite{Gursey:1979tu,Evans:1992az}. Holomorphic quaternion functions seem  too strong a constraint for these
theories, but the construction of a field theory involving quaternion holomorphic functions might the attempted.
On the other hand, M\"obius transformations are metric symmetries for hyperbolic geometries, and they
may be used for defining conformal field theories in such spaces. This has also already been done \cite{Kleban:1993ws}
and makes explicit use of the cross-ratio. A similar study using the restricted cross-ratio discussed here
seems possible. The connection with Heisenberg algebra is also a very interesting 
direction, in order to understand the interpretation in terms of quaternionic quantum mechanics 
of the differential operators of position and momentum. However, there are many other possibilities in this subject, as 
can be seen from the studies that have already been done for with Heisenberg group \cite{Binz:2008ghg}.

\section*{ACKNOWLEDGEMENTS}
Sergio Giardino receives a financial grant under number 206383/2014-2 from the CNPq for his research and is grateful for the hospitality of Professor Paulo 
Vargas Moniz and the Center for Mathematics and Applications of the Beira Interior University. 
%\pagebreak

%%%%%%%%%%%%%%%%%%%%%%%%
%
%
%  BIBLIOGRAPHY
%
%

\bibliographystyle{unsrt} 
\bibliography{bib_qmoebius}

\end{document}